\documentclass[12pt]{article}
\usepackage{epsf}
\usepackage{bm}
\usepackage{cite}
\usepackage{color}
\usepackage{amsmath,amssymb}
\usepackage{graphicx}
\usepackage[colorlinks,citecolor=blue]{hyperref}
\usepackage{comment}
\newcommand{\alm}{a_{lm}}
\newcommand{\talm}{\tilde a_{lm}}
\newcommand{\cl}[1]{C_l^{#1}}

\renewcommand{\thefootnote}{\fnsymbol{footnote}}
\def\thefootnote{\fnsymbol{footnote}}

\makeatletter

\@addtoreset{equation}{section}
\makeatother

\begin{document}

\begin{titlepage}

\flushright{RESCEU-17/23}

\begin{center}

\vskip .45in

{\Large \bf Testing multi-field inflation with LiteBIRD}

\vskip .65in

{\large 
Ryusuke Jinno$^{1}$,
Kazunori Kohri$^{2,3,4,5,6}$,
Takeo~Moroi$^{7,6,4}$, 
Tomo~Takahashi$^{8}$ \vspace{2mm} \\
and
Masashi Hazumi$^{3,4,5,6,9}$ }

\vskip 0.2in

{\em 
$^{1}$Research Center for the Early Universe, The University of Tokyo, \\Tokyo 113-0033, Japan \vspace{2mm} \\
$^{2}$Division of Science, National Astronomical Observatory of Japan (NAOJ), 2-21-1 Osawa, Mitaka, Tokyo 181-8588, Japan
 \vspace{2mm} \\
$^{3}$Institute of Particle and Nuclear Studies (IPNS), High Energy Accelerator Research Organization (KEK), Tsukuba, Ibaraki 305-0801, Japan \vspace{2mm} \\
$^{4}$International Center for Quantum-field Measurement Systems for Studies of the Universe and Particles (QUP), High Energy Accelerator Research Organization (KEK), Tsukuba, Ibaraki 305-0801, Japan \vspace{2mm} \\
$^{5}$The Graduate University for Advanced Studies (SOKENDAI), Miura District, Kanagawa 240-0115, Hayama, Japan \vspace{2mm} \\
$^{6}$Kavli Institute for the Physics and Mathematics of the Universe (Kavli IPMU, WPI), UTIAS, The University of Tokyo, Kashiwa, Chiba 277-8583, Japan \vspace{2mm} \\
$^{7}$Department of Physics, The University of Tokyo, Tokyo 113-0033, Japan \vspace{2mm} \\
$^{8}$Department of Physics, Saga University, Saga 840-8502, Japan \vspace{2mm} \\
$^{9}$Japan Aerospace Exploration Agency (JAXA), Institute of Space and Astronautical Science (ISAS), Sagamihara, Kanagawa 252-5210, Japan
}

\end{center}
\vskip .5in

\begin{abstract}

We investigate expected constraints on the primordial tensor power spectrum from the future cosmic microwave background polarization experiment LiteBIRD as a test of multi-field inflation. We argue that the measurements of the tensor-to-scalar ratio and the tensor spectral index, in combination with the constraints on the scalar spectral index from the Planck observation, are useful in testing multi-field inflation models. We also discuss implications for multi-field inflationary model building.

\end{abstract}

\end{titlepage}

\renewcommand{\thepage}{\arabic{page}}
\setcounter{page}{1}
\renewcommand{\thefootnote}{\#\arabic{footnote}}
\setcounter{footnote}{0}

\section{Introduction} \label{sec:intro}

Inflation is now considered to be a successful paradigm to describe the very early stage of the Universe.
Current cosmological observations of cosmic microwave background (CMB) such as the Planck satellite \cite{Planck:2018jri}, together with other cosmological probes such as baryon acoustic oscillation (BAO) and type Ia supernovae (SNeIa), allow us to accurately probe the properties of primordial density fluctuations on large scales and put severe constrains on the models of inflation. However, the actual mechanism of inflation, or the model of inflation realized in nature has not yet been identified. One of the key quantities in probing inflationary models is the amplitude of the tensor power spectrum, commonly parametrized by the tensor-to-scalar ratio $r$, from which the energy scale of inflation can be extracted. Indeed, in the framework of single-field inflation models, the information on $r$ as well as the spectral index $n_s$ has greatly helped narrow down the models.

However, scalar degrees of freedom are ubiquitous in high energy theories such as supersymmetric models and superstring theories, and thus it is quite conceivable that multiple fields exist during inflation.
In such a multiple field scenario, the inflaton field may just be causing the inflationary expansion, and it may be another field that is responsible for the production of the primordial fluctuations. Examples of such a scenario include the curvaton model~\cite{Enqvist:2001zp,Lyth:2001nq,Moroi:2001ct}, modulated reheating~\cite{Dvali:2003em,Kofman:2003nx}, and so on.
Interestingly, some inflation models excluded by the current data as single-field models become viable in the multi-field framework (see, e.g., \cite{Langlois:2004nn,Moroi:2005kz,Moroi:2005np,Ichikawa:2008iq,Ichikawa:2008ne,Enqvist:2013paa,Morishita:2022bkr} for explicit illustrations).
However, this implies that model degeneracies can easily arise in the prediction of $n_s$ and $r$ in the multi-field framework, which might be an obstacle in pinning down the full theory of inflation. 
Thus, though in much of the literature it is common to investigate only $n_s$ and $r$ for the test of inflationary models, 
in this paper we seek for some other quantities to differentiate models of inflation.

One of the candidates is the spectral index of the tensor mode $n_T$\footnote{
Actually multi-field models generate large non-Gaussianities in some parameter space (see, e.g., \cite{Suyama:2010uj,takahashi:2014bxa} for non-Gaussianities predicted in various models), and hence such models are excluded by the Planck data~\cite{Planck:2019kim}.
However, even in multi-field models, non-Gaussianties can also be small enough to be well within the current observational bound in a broad parameter range, and thus non-Gaussianities are not enough to test the multi-field scenario at least at the current level of constraints.
}.
It is well-known that a consistency relation holds between $r$ and $n_T$ in the single-field inflation with a canonical kinetic term assuming the Bunch-Davies vacuum:
\begin{equation}
\label{eq:consistency_inf}
n_T = -\frac{r}{8}.
\end{equation}
Any deviation from this relation would rule out single-field inflation models, and may point toward multi-field models\footnote{
Other than the multi-field models, the deviation can arise in models with non-standard kinetic term \cite{Garriga:1999vw}, non-Bunch-Davies initial condition \cite{Ashoorioon:2013eia}, and so on.
}.

Although several works have investigated constraints on the tensor spectral index using current cosmological data (e.g., see \cite{Planck:2018jri,Li:2019vlb}), the resulting bounds turned out to be still weak.
Future CMB B-mode satellite experiments such as  LiteBIRD~\cite{Hazumi:2021yqq,LiteBIRD:2022cnt}
can probe not only the tensor-to-scalar ratio but also the tensor spectral index with better sensitivity than current observations, providing more information to probe inflation models.

The aim of this paper is to investigate how multi-field inflation models can be tested with the information on $r$ and $n_T$ in the future LiteBIRD observation, as well as to discuss implications to the construction of such models.
The structure of this paper is as follows. In Section \ref{sec:multi_inf}, we review multi-field inflation models, particularly focusing on spectator field models. In this paper ``multi-field inflation" always refers to models with a spectator field discussed in the section. Then in Section~\ref{sec:method}, we briefly describe our method to investigate the expected constraints on the tensor-to-scalar ratio $r$ and the tensor spectral index $n_T$ in the future LiteBIRD observation. In Section~\ref{sec:constraint}, we show  our results for the constraints and discuss how multi-field inflation models can be tested in LiteBIRD. The final section is devoted to conclusion.

\section{Multi-field inflation} \label{sec:multi_inf}

Here we briefly review the predictions for the scalar and tensor power spectra in multi-field inflation framework, with a particular focus on the so-called spectator field models.
As mentioned in the introduction, ``multi-field inflation" in this paper always refers to models with a spectator field, and thus we use ``multi-field inflation" and ``spectator field model" interchangeably. 
Though large primordial non-Gaussianities can be generated in this kind of scenario in some parameter space, they do not necessarily work as a good probe as they can be small in other parameter range.
In order to illustrate how easily the constraints from non-Gaussianities can be avoided, we also briefly discuss non-Gaussianities in the case of the curvaton model.
For other models of multi-field inflation, we refer the readers to, e.g., \cite{Suyama:2010uj,takahashi:2014bxa}.

\subsection{Scalar and tensor power spectra} 

When there exists a scalar field other than the inflaton during inflation, and if the mass of the former is light enough, it also acquires quantum fluctuations and generate the curvature perturbation.
When the energy density of such a scalar field is negligible during inflation, it is referred to as a ``spectator field."
This kind of scenario includes the curvaton model \cite{Enqvist:2001zp,Lyth:2001nq,Moroi:2001ct}, modulated reheating \cite{Dvali:2003em,Kofman:2003nx} and so on.
Although in the following we mainly discuss the curvaton model, most arguments given below also apply to other spectator field models (see, e.g., \cite{Ichikawa:2008ne,Suyama:2010uj}).

In the simplest curvaton models, the curvature perturbation is assumed to be sourced only from the curvaton field.
However, even in these simplest models, the inflaton also acquires nonzero quantum fluctuations and contribute to the curvature perturbation.
Thus in general the curvature perturbation is a mixture of those derived from the curvaton and the inflaton.
Models that takes both sources into account are called mixed inflaton and curvaton models, and they have been investigated in some detail~\cite{Langlois:2004nn,Moroi:2005kz,Moroi:2005np,Ichikawa:2008iq,Enqvist:2013paa}.
Here we give the formulas for the scalar and tensor power spectra, their spectral indices, and the so-called tensor-to-scalar ratio.

In the mixed inflaton and curvaton (spectator) model, the total curvature (scalar) power spectrum is written as 
\begin{equation}
\label{eq:power_s}
\mathcal{P}_\zeta (k) = \mathcal{P}_\zeta^{(\phi)} (k) + \mathcal{P}_\zeta^{(\sigma)} (k)
= (1+ R_\sigma) \mathcal{P}_\zeta^{(\phi)} (k) , 
\end{equation}
where $ \mathcal{P}_\zeta^{(\phi)} (k)$ and $ \mathcal{P}_\zeta^{(\sigma)} (k)$ are the power spectra generated from the fluctuations of the inflaton $\phi$ and the curvaton (spectator field) $\sigma$, respectively.
In the second equality, we define the ratio
\begin{equation}
\label{eq:R}
R_\sigma \equiv 
 \frac{\mathcal{P}_\zeta^{(\sigma)} }{\mathcal{P}_\zeta^{(\phi)}} \,,
\end{equation}
to express the relative contribution to the power spectrum from the curvaton (spectator field) and the inflaton.
The spectral index $n_s$ and its running $\alpha_s$ of the scalar power spectrum in this mixed model are expressed as  (see, e.g., 
\cite{Kobayashi:2012ba,Sekiguchi:2017cdy}) 
\begin{eqnarray}
\label{eq:ns_mixed}
n_s  -1 &=&  \frac{1}{1+R_\sigma} (  -6 \epsilon + 2 \eta_\phi ) + \frac{R_\sigma}{1+R_\sigma} ( - 2\epsilon +  2 \eta_\sigma) \,,  \\ [12pt]
\label{eq:alpha_mixed}
\alpha_s &=& \frac{1}{1+R_\sigma} (  -24  \epsilon^2  + 16 \epsilon \eta_\phi  - 2 \xi^{(2)}_\phi )  
+ \frac{R_\sigma}{1+R_\sigma} ( -8  \epsilon^2  + 4 \epsilon \eta_\phi  + 4 \epsilon \eta_\sigma  - 2 \xi^{(2)}_\sigma )  \notag \\  [8pt]
&& +  \left( \frac{1}{1+R_\sigma} \right) \left( \frac{R_\sigma}{1+R_\sigma}\right) ( -4  \epsilon  + 2 \eta_\phi  - 2 \eta_\sigma  )^2 
\,,
\end{eqnarray}
where $\epsilon, \eta_\phi, \eta_\sigma, \xi^{(2)}_\phi$ and $\xi^{(2)}_\sigma$ are slow-roll parameters defined as
\begin{eqnarray}
\label{eq:slow-roll}
\epsilon = - \frac{\dot{H}}{H^2} \,,
\quad
\eta_\phi = \frac{W^{\prime\prime}(\phi)}{3H^2} \,,
\quad
\eta_\sigma =  \frac{U^{\prime\prime}(\sigma)}{3H^2} \,,
\quad
\xi^{(2)}_\phi = \frac{W^\prime (\phi) W^{\prime\prime\prime}(\phi)}{(3H^2)^2}  \,,
\quad
\xi^{(2)}_\sigma = \frac{U^\prime (\sigma) U^{\prime\prime\prime}(\sigma)}{(3H^2)^2}  \,. \notag \\
\end{eqnarray}
Here we assume that the potential has a separable form $V(\phi, \sigma) = W(\phi) + U(\sigma)$, where $W(\phi)$ and $U(\sigma)$ are the potentials for the inflaton and the curvaton (spectator field), respectively.
See \cite{Planck:2018jri} for the current constraints on $n_s$ and $\alpha_s$ from the Planck data\footnote{
Although the current constraints on the running parameter $\alpha_s$ are not stringent enough to test inflation models, small-scale observations such as galaxy surveys \cite{Basse:2014qqa,Munoz:2016owz,Li:2018epc}, 21cm fluctuations \cite{Mao:2008ug,Kohri:2013mxa,Munoz:2016owz}, 21cm signal from minihalo \cite{Sekiguchi:2017cdy}, 21cm global signal \cite{Yoshiura:2018zts,Yoshiura:2019zxq}, CMB spectral $\mu$ distortion \cite{Hu:1994bz,Chluba:2012gq,Chluba:2012we,Khatri:2013dha}, galaxy luminosity function \cite{Yoshiura:2020soa}, reionization history \cite{Minoda:2023hvp} can give further information and may allow for more precise measurements of the running in the future.
}.
The above formulas~\eqref{eq:power_s} -- \eqref{eq:slow-roll} are valid for general spectator field models.

To make some explicit calculations, we assume a quadratic potential for the curvaton $\sigma$ as
\begin{equation}
\label{eq:potential_curvaton}
U(\sigma) = \frac12 m_\sigma^2  \sigma^2 \,.
\end{equation}
With this potential, the power spectrum generated from the curvaton is expressed as (see e.g., \cite{Lyth:2002my,Ichikawa:2008iq,Enqvist:2013paa})
\begin{eqnarray}
&&  \mathcal{P}_\zeta^{(\sigma)} = \left( \frac{2 r_{\rm dec}}{3 \sigma_\ast} \right)^2  \left( \frac{H_\ast}{2\pi} \right)^2, 
\end{eqnarray}
where $H_\ast$ and $\sigma_\ast$ are the Hubble parameter and the value of $\sigma$ at the time of horizon crossing during inflation, respectively,
and $r_{\rm dec}$ roughly represents the fraction of the energy density of the curvaton at the time of its decay.
The last quantity can explicitly be written as
\begin{equation}
r_{\rm dec}  = \left. \frac{3 \rho_\sigma}{4\rho_\gamma + 3 \rho_\sigma} \right|_{\rm decay},
\end{equation}
with $\rho_\sigma$ and $\rho_\gamma$ being the energy densities of the curvaton and radiation.

On the other hand, the power spectrum for the inflaton sector can be written as
\begin{eqnarray}
&& \mathcal{P}_\zeta^{(\phi)}  = \frac{1}{2 \epsilon M_{\rm pl}^2} \left( \frac{H_\ast}{2\pi} \right)^2 \,,
\end{eqnarray}
where $\epsilon$ is the slow-roll parameter defined in Eq.~\eqref{eq:slow-roll}.
Though the power spectrum from the inflaton sector depends on the inflaton potential, that dependence can be incorporated into the slow-roll parameter $\epsilon$.

Since the tensor modes are not affected by the existence of the spectator field, their power spectrum has the same expression as the inflaton-only case
\begin{equation}
\label{eq:power_t}
\mathcal{P}_T (k)  = \frac{8}{M_{\rm pl}^2} \left( \frac{H_\ast}{2\pi} \right)^2,
\end{equation}
and similarly the tensor spectral index $n_T$ is given by 
\begin{equation}
\label{eq:nT}
n_T = - 2 \epsilon.
\end{equation}
Since the scalar mode is different from the single-field inflation models, the tensor-to-scalar ratio is modified in the inflaton-spectator mixed models.
From Eqs.~\eqref{eq:power_s} and \eqref{eq:power_t}, the tensor-to-scalar ratio is found to be
\begin{equation}
\label{eq:r}
r = \frac{16 \epsilon}{1+R_\sigma}.
\end{equation}
If we assume the curvaon as the spectator field, we may explicitly write down the expression for $R_\sigma$ with the model parameters as 
\begin{equation}
\label{eq:R_sigma}
R_\sigma = \frac89 \left( \frac{M_{\rm pl} }{\sigma_\ast} \right)^2 r_{\rm dec}^2 \epsilon \,,
\end{equation}
which gives the formula for the tensor-to-scalar ratio in the large $R_\sigma$ limit as
\begin{equation}
r \simeq \frac{16 \epsilon}{R_\sigma} = \frac{18}{r_{\rm dec}^2} \left( \frac{\sigma_\ast}{M_{\rm pl}} \right)^2 \,.
\end{equation}
As briefly discussed in Section~\ref{sec:nonG}, the non-Gaussianity constraints from Planck allow us to set $r_{\rm dec} = 1$.
In this case, when the curvaton gives a dominant contribution to the scalar power spectrum, the initial value of $\sigma$ is given by
\begin{equation}
\sigma_\ast \simeq \sqrt{\frac{r}{18}} M_{\rm pl} \,.
\end{equation}
Therefore, once the tensor-to-scalar ratio is constrained, the value of $\sigma$ can be determined.
Indeed, assuming that the curvaton mainly contribute to the primordial power spectrum, the current constraints on the tensor-to-scalar ratio of Planck + BICEP2018 $r < 0.035~(95~\% {\rm C.L.})$ \cite{Planck:2018jri,BICEP:2021xfz} already imply
\begin{equation}
\sigma_\ast \lesssim 5 \times 10^{-2} \, M_{\rm pl} \,.
\end{equation}
Since $r$ is expected to be detected/severely constrained in LiteBIRD, we expect more information on the curvaton parameter.
Furthermore, if we invoke the arguments suggested by the stochastic formalism~\cite{Starobinsky:1994bd,Starobinsky:1986fx,Enqvist:2012xn,Hardwick:2017fjo}, the typical value of $\sigma$ for the potential of Eq.~\eqref{eq:potential_curvaton} is given by $\sigma_\ast \sim H_\ast^2 / m_\sigma$\footnote{
This is a typical value when the equilibrium distribution is reached for the spectator field.
In some inflation models, it takes long to arrive at the equlibrium distribution~\cite{Hardwick:2017fjo}, and in this case the arguments here do not apply.
}, 
from which we can infer even the mass of the curvaton as $m_\sigma \sim 10^{10} \, (r/0.01)^{1/2} \, {\rm GeV} $.

\begin{figure}[htbp]
  \begin{center}
  \resizebox{90mm}{!}{\includegraphics{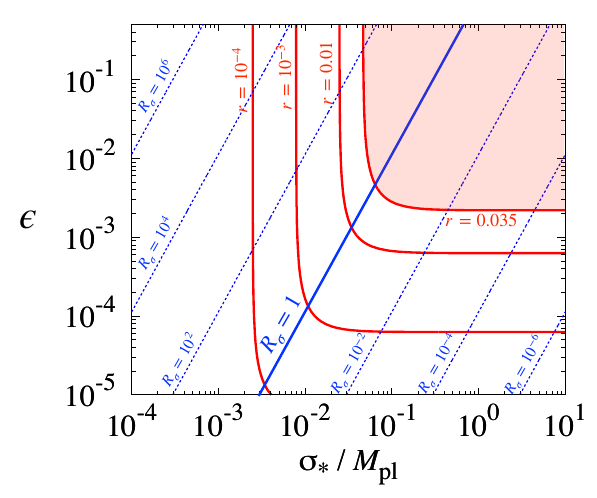}}
  \end{center}
  \caption{Contours of the curvaton-inflaton ratio $R_\sigma$ (blue) and the tensor-to-scalar ratio (red) in the $\sigma_\ast$--$\epsilon$ plane.
  In this figure, the curvaton is assumed to be the spectator.
  From the non-Gaussianity constraints from Planck (see Section~\ref{sec:nonG}), we set $r_{\rm dec} = 1$ in this figure.
  Red shaded region is excluded from the current observations of Planck+BK18.
  }
\end{figure}

Finally we mention the consistency relation in the multi-field inflation models.
By eliminating $\epsilon$ in the expressions of the tensor-to-scalar ratio \eqref{eq:r} and the tensor spectral index \eqref{eq:nT}, we obtain the consistency relation between $r$ and $n_T$:
\begin{equation}
\label{eq:consistency_mixed}
n_T = - (1+R_\sigma)\frac{r}{8}.
\end{equation}
The single-field inflation model corresponds to the limit of $R_\sigma \rightarrow 0$, in which case the relation reduces to the well-known single-field inflation consistency relation $n_T = - r/8$.
If we can observationally probe the consistency relation \eqref{eq:consistency_mixed}, it would be a critical test on the models of the primordial density fluctuations.
Even if we just obtain some constraints on $n_T$, we may still put a bound on $R_\sigma$, which in turn gives the limits on model parameters in multi-field models.
In this respect, the information on $n_T$ would be of great importance in the exploration of the multi-field inflation models.

\subsection{Predictions for $n_s$ and $r$ in the multi-field models} 

As already discussed in several works~\cite{Langlois:2004nn,Moroi:2005kz,Moroi:2005np,Ichikawa:2008iq,Ichikawa:2008ne,Enqvist:2013paa}, in the multi-field framework some single-field inflation models can revive even if they are excluded by the measurements of the spectral index~$n_s$ and the tensor-to-scalar ratio~$r$ such as Planck~\cite{Planck:2018jri} and BICEP~\cite{BICEP:2021xfz}.
This occurs because of the modifications of the predictions of $n_s$ and $r$, in particular because of the suppression of $r$ in the multi-field models, as seen from the factor $R_\sigma$ in the denominator of Eq.~\eqref{eq:r}.
For a light spectator field $|\eta_\sigma | \ll 1$, the spectral index (\ref{eq:ns_mixed}) in the $R_\sigma \to \infty$ limit reduces to
\begin{equation}
\label{eq:ns_spectator_limit}
n_s - 1 = -2 \epsilon \,.
\end{equation}
Thus only the $\epsilon$ parameter determines $n_s$ in the case of a light spectator.

\begin{figure}[htbp]
  \begin{center}
  \resizebox{150mm}{!}{\includegraphics{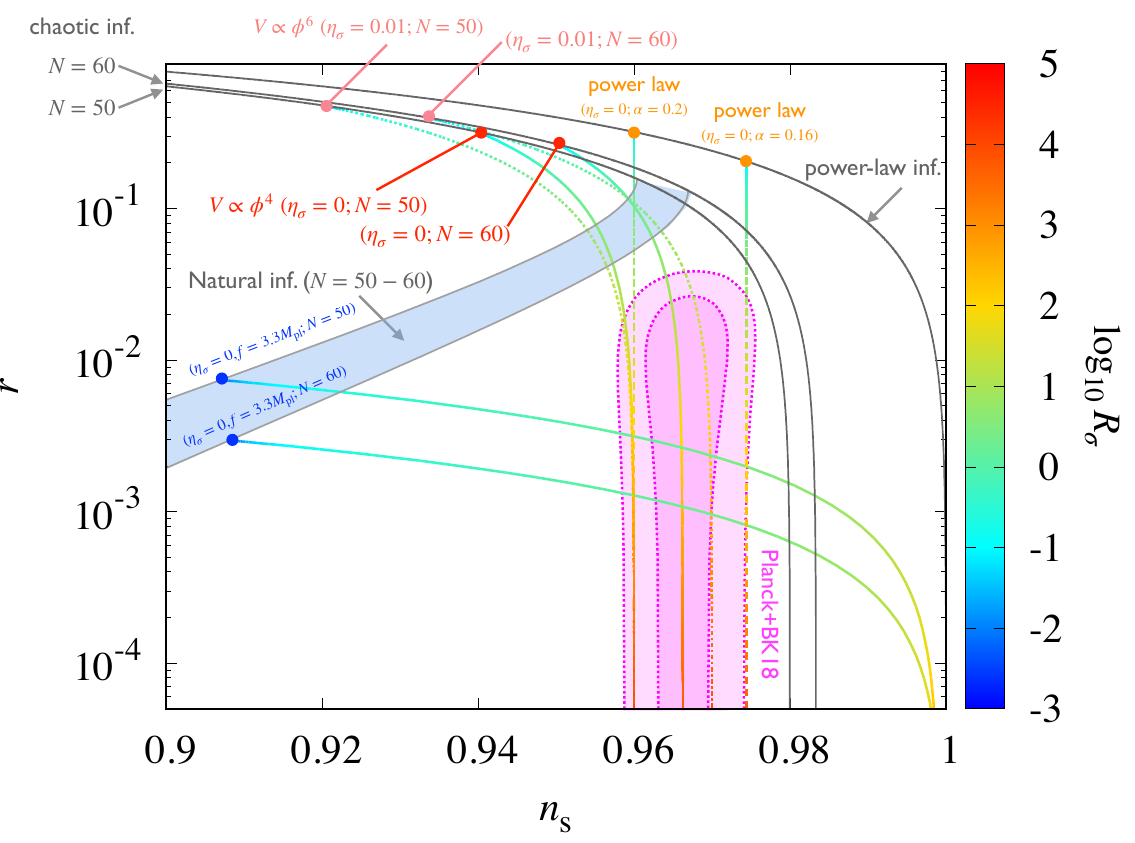}}
  \end{center}
  \caption{\label{fig:ns_r_spectator} 
  Predictions of spectator field models in the $n_s$-$r$ plane.
  We show the cases of chaotic inflation $W \propto \phi^n$ with $n=4$ and $6$, natural inflation $ W \propto  1 - \cos (\phi / f) $ and power-law inflation $ W \propto \exp (- \alpha \phi /M_{\rm pl})$.
  For each inflaton model, the circles represent the predictions of the single-field case, while
  the lines starting from these circles correspond to the predictions of spectator field models.
  The value of $R_\sigma$ increases along each line, i.e., more contribution from the spectator field is added to the scalar power spectrum, as indicated by the color.
  For reference, we also depict the 1$\sigma$ and 2$\sigma$ allowed regions from Planck+BICEP18~\cite{Planck:2018jri,BICEP:2021xfz} (dark and light magenta regions).
  As seen from the behavior of the lines, even if a model is excluded as a single-field inflation $R_\sigma = 0$, it can be revived when the contribution from the spectator field is added.
  }
  \end{figure}

To demonstrate these properties explicitly, we depict the predictions of multi-field models in the $n_s$-$r$ plane in Fig.~\ref{fig:ns_r_spectator}.
For the inflaton models we assume the chaotic inflation $V \propto \phi^n$, natural inflation $ V \propto  1 - \cos (\phi / f) $, and power-law inflation $ V \propto \exp (- \alpha \phi /M_{\rm pl})$.
In this figure we show the predictions for these inflaton models with different values of $R_\sigma$, the contribution to the curvature perturbation from the spectator field (\ref{eq:R}), in different colors.
For reference, we also show the constraints from Planck+BICEP (denoted as Planck+BK18 in the figure) \cite{Planck:2018jri,BICEP:2021xfz}. 

While the chaotic inflation with $n = 4$ is already ruled out as a single-field model (corresponding to the red circles representing the single-field limit for $N = 50$ and $60$), in the multi-field framework it becomes viable for a light spectator $\eta_\sigma \simeq 0$ because of the suppression of the tensor-to-scalar ratio.
Even the case with $n=6$, which predicts a too large value of the tensor-to-scalar ratio to match the current constraints as a single-field model, comes back to life by adding the contribution from a slightly heavy spectator field ($\eta_\sigma = 0.01$).

The natural inflation model is also ruled out as a single-field model by the Planck+BK18 constraints, particularly because it predicts too much red-tilted spectrum coming from the negatively large value of $\eta_\phi$ and very small $\epsilon$ for $f \sim M_{\rm pl}$.
In the blue circles in Fig.~\ref{fig:ns_r_spectator}, we assume $f = 3.3 M_{\rm pl}$, in which case the spectral index $n_s \simeq 0.91$ is far below the current lower bound on $n_s$.
However, once a spectator gives a dominant contribution to the scalar power spectrum, the spectral index approaches to  $n_s \rightarrow 1 - 2 \epsilon$ as $R_\sigma $ gets larger, as shown in the colored lines in the figure.
As a result, at some value of $R_\sigma$ the scalar spectral index becomes consistent with the current constraints.

We also show the predictions of the power-law inflation model $V \propto \exp (- \alpha \phi /M_{\rm pl})$ with $\alpha$ being a model parameter.
In this model, the slow-roll parameters are given by $\epsilon = \alpha^2/2$ and $\eta_\phi = \alpha^2$ and they do not evolve in time during inflation.
Hence the end of inflation should be realized by some mechanism such as a water-fall field rather than by the violation of slow roll.
In this model, $n_s$ and $r$ are related by $r = -8 (n_s -1)$ as shown in the figure, which cannot satisfy the observational constraints by Planck+BK18.
For example, when $\alpha = 0.16$ and $0.2$, the spectral index becomes $n_s = 0.974$ and $0.96$, respectively, which falls within the observationally allowed region, while the tensor-to-scalar ratio $r= 0.2$ and $0.32$ are far outside the allowed region.
However, as in the case of the chaotic inflation, the tensor-to-scalar ratio gets suppressed in the multi-field framework and the power-law inflation become viable.

As shown in these examples, some inflation models excluded as single-field models can still satisfy observational constraints such as Planck+BK18 in the framework of multi-field inflation model.
On the other hand, there exist some single-field inflation models such as the $R^2$ model~\cite{Starobinsky:1980te} and the Higgs inflation~\cite{Cervantes-Cota:1995ehs,Bezrukov:2007ep} that predict $n_s$ and $r$ well inside the observationally allowed region without any spectator\footnote{
Indeed, by introducing a non-minimal coupling to gravity, the predictions for $n_s$ and $r$ are modified to make some inflation models viable.
See, e.g., \cite{Linde:2011nh,Boubekeur:2015xza,Tenkanen:2017jih,Ferreira:2018nav,Antoniadis:2018yfq,Takahashi:2018brt,Shokri:2019rfi,Takahashi:2020car,Reyimuaji:2020goi,Cheong:2021kyc,Kodama:2021yrm} for works along this line.
}.
Thus it would be worth investigating whether one may differentiate single- and multi-field models, which we argue in this paper,  is possible to some extent from the observations of the tensor power spectrum by the future LiteBIRD experiment.
Before discussing this main subject, we briefly mention non-Gaussianity in multi-field models.

\subsection{Non-Gaussianities \label{sec:nonG}} 

Here we briefly discuss non-Gaussianities in the curvaton model and argue that they do not severely constrain multi-field models at the current level of observational bounds\footnote{
We should also note that future observations of galaxy surveys, 21cm fluctuations and so on could improve the bound \cite{Yamauchi:2014ioa,Munoz:2015eqa,Yamauchi:2015mja,Sekiguchi:2018kqe}.
}.

Non-Gaussianities can be quantified by higher order statistics such as bi- and tri-spectra.
Since the trispectrum constraints are currently not so strong, here we only discuss the bispectrum.
Although several functional forms for the bispectrum have been discussed in the literature, the spectator field models considered in this paper generate the so-called local-type one, in which the bispectrum $B_\zeta (k_1, k_2, k_3)$ can be written as 
\begin{equation}
B_\zeta (k_1, k_2, k_3) = \frac65 f_{\rm NL} \left[ 
P_\zeta (k_1) P_\zeta (k_2) + P_\zeta (k_2) P_\zeta (k_3) + P_\zeta (k_3) P_\zeta (k_1)  
\right] \,.
\end{equation}
Here $f_{\rm NL}$ is the non-linearity parameters characterizing the size of the bispectrum.
It is commonly taken to be constant, and we assume this to be the case in this paper, but we note that it can be scale-dependent in some cases~\cite{Byrnes:2009pe,Byrnes:2010ft,Byrnes:2010xd,Huang:2010cy,Huang:2010es,Byrnes:2011gh,Huang:2011py,Kobayashi:2012ba,Byrnes:2015asa}.

In the curvaton model, the non-linearity parameter $f_{\rm NL}$ is given by\footnote{
The formula below is applicable to the cases where the potential of the curvaton is given by \eqref{eq:potential_curvaton}.
When the potential deviates from the quadratic form, the prediction for non-Gaussianities gets modified~\cite{Sasaki:2006kq,Enqvist:2008gk,Enqvist:2009eq,Enqvist:2009ww,Suyama:2010uj}.
} \cite{Bartolo:2003jx,Lyth:2005fi,Lyth:2005du,Sasaki:2006kq}
\begin{equation}
f_{\rm NL} = \frac{5}{4 r_{\rm dec}} - \frac53 - \frac{5}{6} r_{\rm dec} \,.
\end{equation}
Stringent bounds on the non-linearity parameters are obtained by the Planck data, and the constraints on the local type is $f_{\rm NL} = -0.9 \pm 5.1 \,\, (95 \% \, {\rm CL})$ \cite{Planck:2019kim}, which translates to the limit $r_{\rm dec} > 0.21  \,\, (95 \% \, {\rm CL})$~\cite{Planck:2019kim}.
Indeed, $r_{\rm dec}$ is roughly given by the fundamental model parameters as 
\begin{equation}
r_{\rm dec} \sim \min \left[ 1, \left( \frac{\sigma_\ast}{M_{\rm pl}} \right)^2 \sqrt{\frac{m_\sigma}{\Gamma_\sigma}} \right] \,,
\end{equation}
Here the model parameters are the mass $m_\sigma$, decay rate $\Gamma_\sigma$, and the initial value $\sigma_\ast$ of the curvaton.
One sees that $r_{\rm dec}$ can be of order unity in a broad parameter range and that the constraints from Planck can easily be satisfied.
Thus it is difficult to comprehensively test multi-field inflation even with non-Gaussianties.
However, as we argue in Section~\ref{sec:constraint}, the tensor-to-scalar ratio and the tensor spectral index may serve as a crucial test for multi-field inflation models.

\section{Analysis method} \label{sec:method}

In this section we briefly describe our methods to investigate the expected constraints on the tensor-to-scalar ratio $r$ and the tensor spectral index $n_T$ from the future CMB B-mode polarization experiment of 
LiteBIRD \cite{Hazumi:2021yqq,LiteBIRD:2022cnt}.
Although we sometimes take account of the Planck constraints on the spectral index~\cite{Planck:2018jri}, we basically use the B-mode polarization from LiteBIRD alone, since the aim of this paper is to show the usefulness of the observables from the tensor modes for model selection.  For further details of the analysis method, we refer the readers to~\cite{Verde:2005ff,Baumann:2008aq,Creminelli:2015oda,Errard:2015cxa,Barron:2017kuo,Oyama:2015gma}.
Our treatment is simpler than the more rigorous methods in e.g., 
\cite{LiteBIRD:2022cnt,Stompor:2016hhw,Errard:2018ctl,Verges:2020xug,LiteBIRD:2023iei}, 
but as we will mention later, we obtain almost the same constraint on $r$ as the one given in \cite{LiteBIRD:2022cnt} by tuning the parameter representing the fraction of the residual foreground. Therefore 
we believe our simple treatment would give a reasonable constraints on $r$ and $n_T$ which can be attainable from LiteBIRD.

\subsubsection*{Likelihood function and effective $\chi^2$}

The likelihood function for the CMB map is given by~\cite{Bond:1998qg}
\begin{equation}
{\cal L} \propto \prod_{l, m}  \frac{1}{\sqrt{\det {\bm C}_l}}  \exp \left[ -\frac12 \tilde{\bm a}_{l m}^\dagger {\bm C}_l^{-1} \tilde{\bm a}_{l m} \right] \,.
\end{equation}
In general, $\tilde{\bm a}_{lm}$ is a vector constructed from the coefficients of the spherical harmonic expansion of the temperature, $E$ and $B$ polarizations, and the lensing-induced deflection maps from mock data.
However, below we only consider $B$-mode as stated above, and hence $\tilde{\bm a}_{lm}$ just corresponds to $a_{lm}^B$ in our analysis.
The matrix ${\bm C}_l$ is the covariance matrix for a given cosmological model, but in the following we just take it to be $C_l^{BB} = \left\langle \alm^{B\ast} \alm^B \right\rangle$.
Since we only consider the $B$-mode, the effective $\chi_{\rm eff}^2$ is calculated as 
\begin{equation}
\label{eq:chi2_eff}
\chi_{\rm eff}^2 = -2 \ln {\cal L} = \sum_{l,m} \ln ( \det C_l^{BB} )  + \tilde{a}_{lm}^{B \dagger} (C_l^{BB})^{-1} \tilde{a}_{lm}^B \,.
\end{equation}
By fixing the normalization of ${\cal L}$ such that ${\cal L}$ becomes unity (i.e., $\chi^2_{\rm eff} = 0$) for the most likely hypothesis, we write  $\chi_{\rm eff}^2$ as
\begin{eqnarray}
\label{eq:chi2_eff_B}
&& \chi_{\rm eff}^2 = \sum_l (2l+1)  \left[ 
\ln \left( \frac{\cl{BB}}{\tilde{C}_l^{BB}} \right) + \frac{\tilde{C}_l^{BB}}{\cl{BB}} - 1 
\right] \,,
\end{eqnarray}
where we used 
\begin{equation}
\label{eq:C_l_data}
\sum_m \talm^{B\ast} \talm^{B} = (2 l +1) \tilde{C}_l^{BB} \,.
\end{equation}
Notice that $\tilde{C}_l^{BB}$ corresponds to the angular power spectrum for the fiducial model, while $C_l^{XY}$ represents the one for each hypothesized model.
The $\cl{}$'s include the signal and noise power spectra, and hence $C_l^{BB} = C_l^{BB {\rm (sig)}} + N_l^{BB}$ where $C_l^{BB {\rm (sig)}}$ is the signal power spectrum which is given by the fiducial power spectrum for $\cl{BB}$.
Furthermore, in addition to the instrumental noise, the contributions from the foreground and delensing can effectively be included in the noise power spectrum, and thus $N_l^{BB}$ is decomposed as $N_l^{BB} = N_{l {\rm (inst) }}^{BB} + F_l ^{BB} + C_l^{BB, {\rm res}}$.
Here $ N_{l {\rm (inst) }}^{BB}, F_l^{BB}$, and $C_l^{BB, {\rm res}}$ are respectively the power spectra for the instrumental noise, foreground, and delensing, which we briefly describe below.

\subsubsection*{Instrumental noise}

Assuming uncorrelated noise for different modes, the noise power spectrum $N_{{\rm (inst)} l}^{BB}$ for the polarization modes is given by (e.g., \cite{Verde:2005ff,Baumann:2008aq,Creminelli:2015oda,Errard:2015cxa,Barron:2017kuo,Oyama:2015gma}) 
\begin{equation}
\label{eq:inst_noise_N}
N_{l{\rm (inst)}}^{BB} (\nu_i) = (\Delta_B (\nu_i))^2  \exp \left[ l(l+1) \frac{\theta_{\rm FWHM}^2}{8 \ln 2}\right], 
\end{equation}
where $\Delta_B (\nu_i)$ is the instrumental noise for the frequency channel $\nu_i$ in units of $\mu {\rm K} \cdot {\rm radian}$, and $ \theta_{\rm FWHM}$ is the full-width at half-maximum beam size.
When multiple frequency channels are available as is the case for LiteBIRD, we also need to take account of the foreground power spectra together.
We give the total noise power spectrum after briefly discussing the foreground power in the following.

In Table~\ref{tab:spec_LiteBIRD}, we show the specification of the LiteBIRD experiment \cite{LiteBIRD:2022cnt}, which we adopt in our analysis.
We omit the lowest and highest frequency bands for the sake of foreground removal.
In several frequency bands, multiple telescopes and/or  multiple detectors on a single telescope observe the same band with the same center frequency, but with different sensitivities and beam sizes.
In such a case, we regard those bands as separate ones and treat them independently, then combine them in the manner as will be given below.

\begin{table}[hb]
  \centering
  \begin{tabular}{c|cccc}
\hline
Telescope & Center frequency & Polarization Sensitivity $(\Delta_B)$ & Beam Size $(\theta_{\rm FWHM})$ \\
& (GHz) &  ($\mu$K$\cdot$arcmin) & (arcmin)  \\ \hline
LFT & 40 & 37.42 & 70.5  \\  \hline
LFT & 50 & 33.46 & 58.5 \\  \hline
LFT & 60 & 21.31 & 51.1 \\  \hline 
LFT & 68 & 19.91 & 41.6 \\
 & 68 & 31.77 & 47.1 \\  \hline
LFT & 78 &  15.55 & 36.9 \\
 & 78 &  19.13& 43.8\\  \hline
LFT & 89 & 12.28 & 33.0 \\
 & 89 & 28.77 & 41.5 \\ \hline
LFT & 100 & 10.34 &  30.2\\
MFT & 100 & 8.48 & 37.8 \\  \hline
LFT & 119 & 7.69 & 26.3 \\
MFT & 119 & 5.70 & 33.6 \\  \hline
LFT & 140 & 7.25 & 23.7 \\
MFT & 140 & 6.38 & 30.8 \\ \hline
MFT & 166 & 5.57 & 28.9 \\ \hline
MFT & 195 & 7.05 & 28.0 \\
HFT & 195 & 10.50 & 28.6 \\ \hline
HFT & 235 & 10.79 & 24.7 \\ \hline
HFT & 280 & 13.80 & 22.5 \\ \hline
HFT &337 & 21.95 & 20.9 \\ \hline
HFT & 402 & 47.45 & 17.9 \\
\hline
\end{tabular} 
  \caption{\label{tab:spec_LiteBIRD} Specification of LiteBIRD \cite{LiteBIRD:2022cnt}. The lowest (40 GHz) and highest (402 GHz) hands are assumed to be used for foreground removal. 
  We use all the other frequency bands in our analysis.}
\end{table}

\subsubsection*{Foreground}

In the foreground of the polarization maps, there exist two types of contributions: synchrotron emission and thermal emission from dust. The residual foreground power spectrum $F^{BB}_{l} (\nu_i)$ for a single frequency channel~$\nu_i$ is given by 
\begin{equation}
\label{eq:foreground_F}
F_{l}^{BB}  (\nu_i) = \left(C_ l^{BB, {\rm syn}} (\nu_i) + C_l^{BB, {\rm dust}} (\nu_i)  \right) \sigma_{BB, {\rm FG}}^2  + N_{l}^{{BB, \rm FG}} (\nu_i),
\end{equation}
where $C_{ l}^{BB, {\rm syn}} (\nu_i) $ and $C_{l}^{BB, {\rm dust}} (\nu_i)$  are contributions from synchrotron emission and thermal emission from dust.
In Appendix~\ref{app:foreground} we summarize the explicit forms of these power spectra, as well as the one for the noise power spectrum of the foreground template map $N_{l}^{{BB, \rm FG}} (\nu_i)$.
Also, $\sigma_{BB, {\rm FG}}$ represents the fraction of the residual foreground for the $B$-mode. We take $\sigma_{BB, {\rm FG}} = 0.2$ in the analysis such that our treatment gives the 1$\sigma$ uncertainty of the tensor-to-scalar ratio $\Delta r = 5 \times 10^{-4}$, which is consistent with the value obtained in the recent analysis \cite{LiteBIRD:2022cnt} 
for the fiducial value $r=0$ 
with fixed $n_t$ adopting a more rigorous treatment of the foreground.
This identification of the value of $\sigma_{BB, {\rm FG}}$ allows us to provide almost the same constraint on the tensor mode as a more rigorous analysis.

After specifying $C_ l^{BB, {\rm syn}} (\nu_i), C_l^{BB, {\rm dust}} (\nu_i)$, and $N_{l}^{{BB, \rm FG}} (\nu_i)$, the foreground power spectrum $F^{BB}_{l} (\nu_i)$ for a single frequency channel $i$ can be calculated.
Defining the effective noise power spectrum for a frequency channel $i$ with the instrumental noise~\eqref{eq:inst_noise_N} and the foreground~\eqref{eq:foreground_F} as
\begin{equation}
N_{{\rm eff}, l}^{BB} (\nu_i) = N^{BB}_{l {\rm (inst)}} (\nu_i) + F^{BB}_{l} (\nu_i) \,,
\end{equation}
we obtain the total noise power spectrum
\begin{equation}
\label{eq:N_eff_tot_B}
\left( N_{{\rm eff},l}^{BB {\rm (tot)}}  \right)^{-1} = \sum_{i =1}^{N_{\rm ch}} \left[ \frac{1}{  N_{{\rm eff},l}^{BB} (\nu_i) } \right] \,,
\end{equation} 
where $N_{\rm ch}$ is the number of the channels. In addition to the above total noise power spectrum including the foreground, we also take account of the residual lensed $B$-mode as part of the noise power spectrum, which is discussed below.

\subsubsection*{Delensing}

To evaluate the lensed $B$-mode generated from the $E$-mode, we follow \cite{Smith:2010gu}, in which delensing has been performed using CMB polarization only.
For other methods such as the one using the  large-scale structure and cosmic infrared background,  
see e.g., \cite{Smith:2010gu,Sherwin:2015baa,BaleatoLizancos:2021owo}.

The $B$-mode arising from the leakage of the $E$-mode through lensing is evaluated as~\cite{Smith:2010gu}
\begin{equation}
\label{eq:C_lens}
C_{l_1}^{BB, {\rm lensed}} =
\frac{1}{2l_1+1} \sum_{l_2, l} \left| f_{ l_1 l_2 l}^{EB}  \right|^2
C_{l_2}^{EE} C_l^{\phi\phi} \,,
\end{equation}
where $C_{l}^{EE}$ and $\cl{\phi\phi}$ are the power spectra for the polarization $E$-mode and the lensing potential $\phi$, respectively.
Here $ f_{l_1  l_2  l}^{EB} $ is a geometric factor\footnote{
The explicit expression for $ f_{l_1  l_2  l}^{EB} $ is
\begin{equation}
\label{eq:f_lll}
 f_{ l_1  l_2 l}^{EB} = \frac{1}{2 i} \left( F_{l_1  l_2 l } ^{-2}  - F_{l_1  l_2 l } ^{2}  \right) \,, 
\end{equation}
where 
\begin{equation}
F^s_{l_1, l_2, l_3} 
= 
\left[  - l_1 (l_1+1) + l_2 (l_2 +1) + l_3 (l_3+1) \right] 
\sqrt{\frac{(2l_1+ 1) (2l_2+ 1) (2l_3+ 1) }{16 \pi}} 
\begin{pmatrix}
l_1& l_2 & l_3 \\
-s & s & 0
\end{pmatrix}
\,,
\end{equation}
with $
\begin{pmatrix}
l_1& l_2 & l_3 \\
-s & s & 0
\end{pmatrix}
$ being the Wigner 3-$j$ symbol.
}.
The estimated lensed $B$-mode constructed from $E$-mode and lensing potential power spectra is given by 
\begin{equation}
\label{eq:C_lens_est}
C_{l_1}^{BB, {\rm est}} =
\frac{1}{2l_1+1} \sum_{l_2, l} \left| f_{ l_1 l_2 l}^{EB}  \right|^2
\frac{(C_{l_1}^{EE})^2}{C_{l_1}^{EE} + N_{l_1}^{EE}}  \frac{(C_{l_2}^{\phi\phi})^2}{ C_{l_2}^{\phi\phi} + N_{l_2}^{\phi\phi}} \,,
\end{equation}
where the noise power spectrum for the $E$-mode $N_l^{EE}$ is given by the same expression as that for the $B$-mode \eqref{eq:inst_noise_N}.
The noise power spectrum for the lensing potential $N_l^{\phi\phi}$ is given by \cite{Smith:2010gu}
\begin{equation}
N_{l}^{\phi\phi} = 
\left[ 
\frac{1}{2l+1} \sum_{l_1, l_2} \left| f_{ l_1 l_2 l}^{EB}  \right|^2
\frac{1}{C_{l_1}^{BB} + N_{l_1}^{BB}}  \frac{(C_{l_2}^{EE})^2}{ C_{l_2}^{EE} + N_{l_2}^{EE}} 
\right]^{-1} \,.
\end{equation}
Using Eqs.~\eqref{eq:C_lens} and \eqref{eq:C_lens_est}, we may estimate the residual lensed $B$-mode after delensing as \cite{Smith:2010gu}
\begin{equation}
\label{eq:lens_residual_B}
C_{l_1}^{BB, {\rm res}} = C_{l_1}^{BB, {\rm lensed}} - C_{l_1}^{BB, {\rm est}} 
=
\frac{1}{2l_1+1} \sum_{l_2, l} \left| f_{ l_1 l_2 l}^{EB}  \right|^2
\left[ 
C_{l_2}^{EE} C_l^{\phi\phi}
- 
\frac{(C_{l_1}^{EE})^2}{C_{l_1}^{EE} + N_{l_1}^{EE}}  \frac{(C_{l_2}^{\phi\phi})^2}{ C_{l_2}^{\phi\phi} + N_{l_2}^{\phi\phi}} 
\right] \,.
\end{equation}
In our actual implementation, we use real-space expressions to obtain $C_{l_1}^{BB, {\rm res}} $ and $N_{l}^{\phi\phi}$ to make numerical computations faster~\cite{Smith:2010gu}.

Now we have the formula to evaluate the residual lensed B mode.
The total effective noise power spectrum is calculated as
\begin{equation}
N_l^{BB} = N_{{\rm eff}, l}^{BB \rm (tot)} + C_{l}^{BB, {\rm res}} \,,
\end{equation}
where $N_{{\rm eff}, l}^{BB \rm (tot)}$ and $C_{l}^{BB, {\rm res}}$ are respectively given in Eqs.~\eqref{eq:N_eff_tot_B} and \eqref{eq:lens_residual_B}.
With this total effective noise, the value of $\chi^2_{\rm eff}$ is calculated by substituting $ C_l^{BB} + N_l^{BB}$ and $\tilde{C}_l^{BB} + \tilde{N}_l^{BB}$ for $C_l^{BB}$ and $\tilde{C}_l^{BB}$ in Eq.~\eqref{eq:chi2_eff_B}, respectively.
Then we perform a $\chi^2$-based analysis to obtain the expected constraints from LiteBIRD.
In contrast to many other studies, we do not adopt the Fisher matrix analysis because it does not give accurate results especially when we explore the parameter regions around the very edge of the observable boundary.
We highlight this point in the next section.

\section{Expected constraints from LiteBIRD as a probe of multi-field inflation} \label{sec:constraint}

In this section we present the results for the expected constraints on the tensor-to-scalar ratio $r$ and the tensor spectral index $n_T$ from the LiteBIRD experiment. We also show the constraints on the model parameters for several multi-field benchmark models. Below we take the fiducial model to be (i) single-field inflation and (ii) multi-field inflation. Results for these two cases are presented separately in order.

\begin{figure}[htbp]
  \begin{center}
  \resizebox{170mm}{!}{\includegraphics{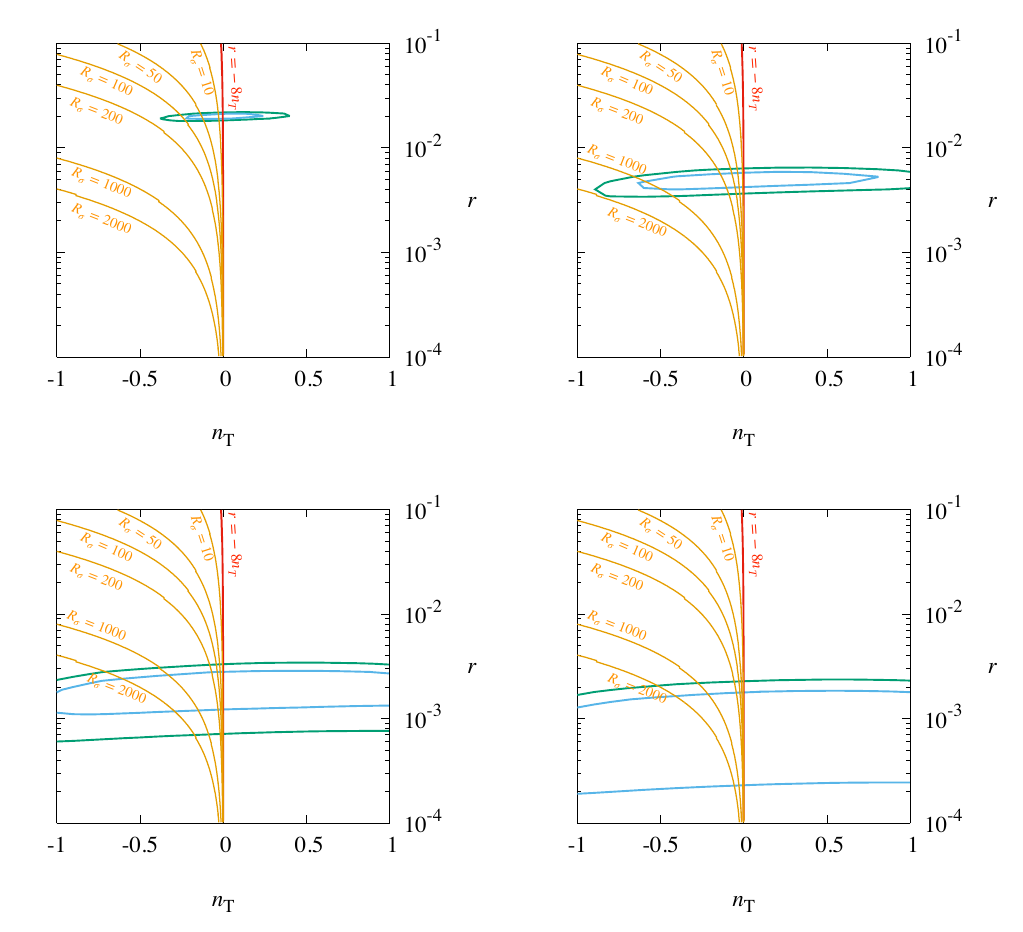}}
  \end{center}
  \caption{\label{fig:nt_r_single}
  1$\sigma$ and 2$\sigma$ expected constraints on the $n_T$--$r$ plane from LiteBIRD.
  The fiducial values are taken as $(r,n_T) = (0.02, -0.0025)$ (top left),  $(0.005, -0.000625)$ (top right), $(0.002, -0.00025)$ (bottom left), and $(0.001, -0.000125)$ (bottom right), where  the consistency relation between $r$ and $n_T$ for single-field inflation models is satisfied.
  The relations between $r$ and $n_T$ for multi-field cases  \eqref{eq:consistency_mixed} are also shown for $R_\sigma=10, 50, 100, 200, 1000$, and $2000$.
  }
\end{figure}

\subsection{Case (i): single-field fiducial points}

In Fig.~\ref{fig:nt_r_single}, we show the expected constraints in the $n_T$--$r$ plane from LiteBIRD assuming the fiducial values of $(r,n_T) = (0.001, -0.000125)$, $(0.002, -0.00025)$, $(0.005, -0.000625)$, and $(0.02, -0.0025)$.
These fiducial points satisfy the consistency relation \eqref{eq:consistency_inf} for the single-field inflation between $r$ and $n_T$.
We take the reference scale as $k_{\rm ref} = 0.01~{\rm Mpc}^{-1}$ at which the tensor-to-scalar ratio and the tensor spectral index are defined.
With this choice, the constraints on $r$ and $n_T$ are almost uncorrelated over the range of the scale probed by LiteBIRD~\cite{Huang:2015gca}.
In Fig.~\ref{fig:nt_r_single}, we also show as yellow lines the predictions of the multi-field inflation models in the $r$--$n_T$ plane for given values of $R_\sigma$ (see Eq.~\eqref{eq:consistency_mixed}).
As seen from Eq.~\eqref{eq:r}, there exists a degeneracy between $\epsilon$ and $R_\sigma$ for a given value of the tensor-to-scalar ratio $r$.
However, once the information on $n_T$ is obtained, even if it is relatively loose, it helps estimate the value of $\epsilon$, and thus the value of $R_\sigma$ can be constrained without specifying the value of $\epsilon$.
Therefore, although checking the single-field consistency relation $n_T = - r/ 8$ would be challenging in LiteBIRD, the information of $n_T$ would still be very useful in testing multi-field models.

\begin{figure}[th]
  \begin{center}
  \resizebox{80mm}{!}{\includegraphics{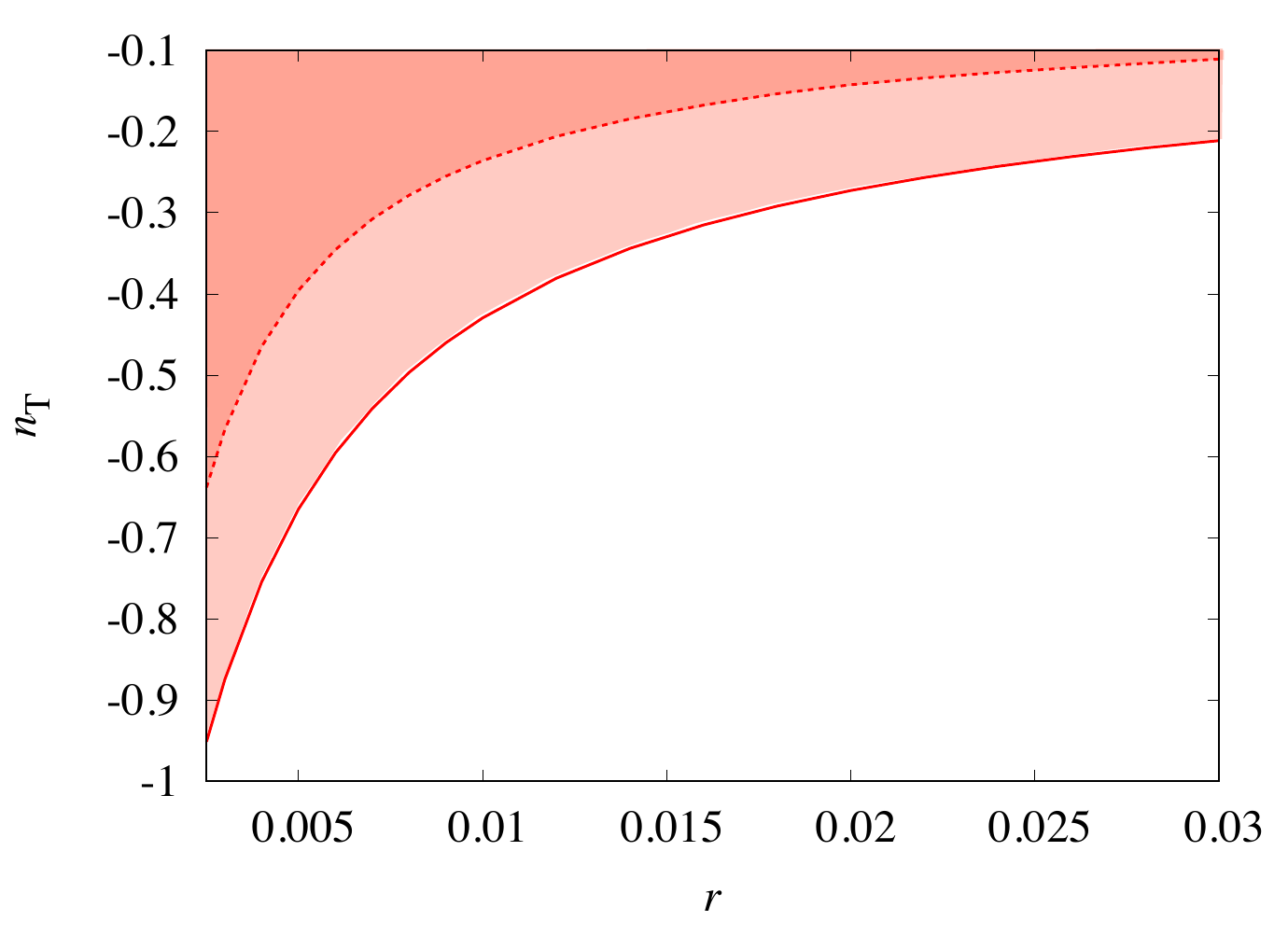}}\hspace{3mm}\resizebox{80mm}{!}{\includegraphics{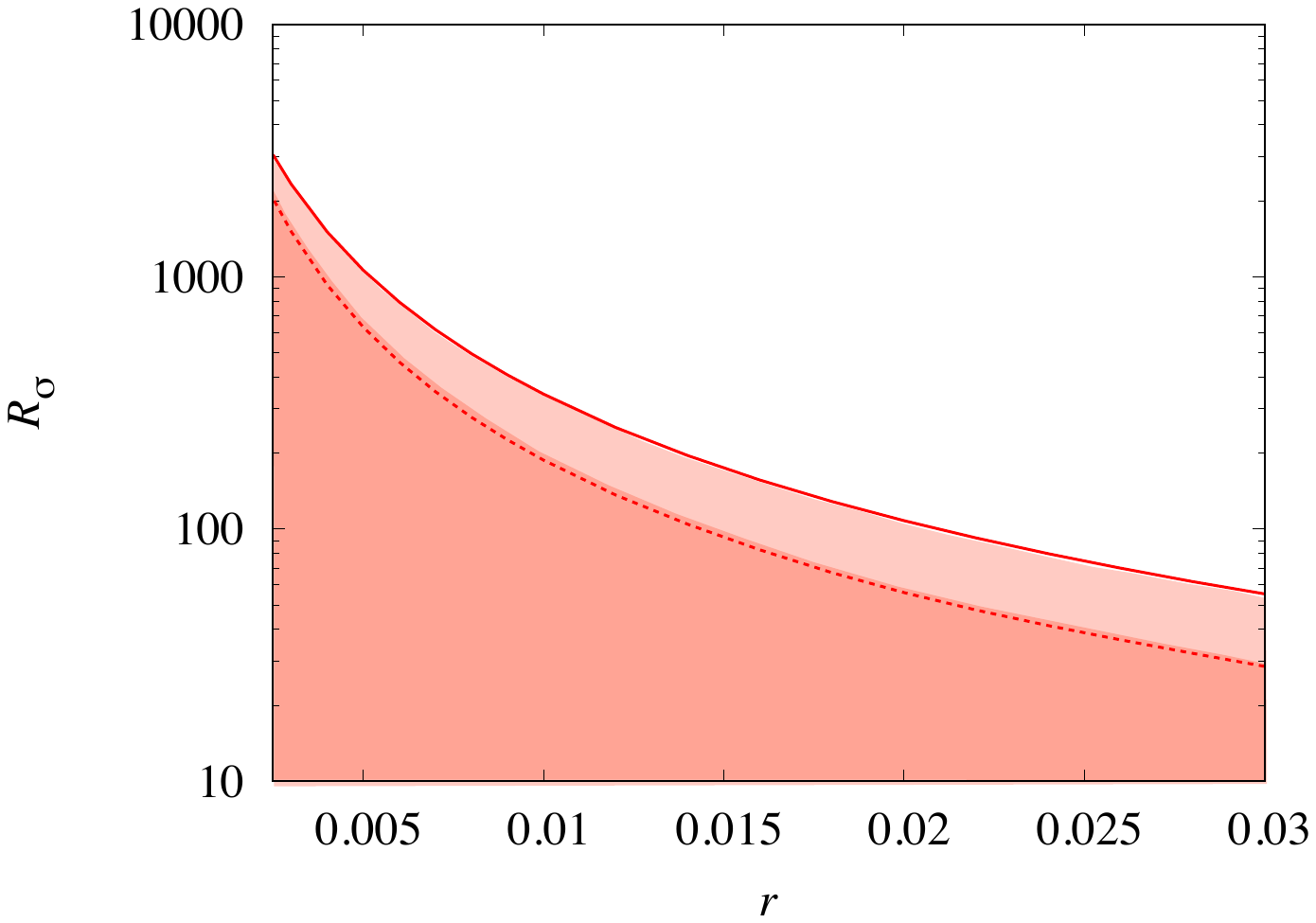}}
  \end{center}
  \caption{ \label{fig:nT_limit}
  1$\sigma$ (dark red) and 2$\sigma$ (light red) expected constraints on $n_T$ (left panel) and $R_\sigma$ (right panel) for fixed values of $r$ from LiteBIRD.
  }
\end{figure}

In Fig.~\ref{fig:nt_r_single}, the constraints on $n_T$ get weaker as the fiducial value of $r$ becomes smaller.
To see this more clearly, we plot the expected constraints on $n_T$ from LiteBIRD for fixed values of $r$ in the left panel of Fig.~\ref{fig:nT_limit}.
Since $n_T$ is always negative in our framework of multi-field inflation, we show only the negative region of $n_T$.
When $r \sim 0.02$, the tensor spectral index is constrained to be $n_T > -0.2$ at 1$\sigma$ level.
On the other hand, when $r \lesssim 0.002$, it cannot be severely constrained and even $n_T \simeq -1$ cannot be excluded. 

Since from Eq.~\eqref{eq:consistency_mixed} $R_\sigma$ is expressed with $r$ and $n_T$ as 
\begin{equation}
R_\sigma = - \frac{8n_T}{r} - 1 \,,
\end{equation}
the constraints on $n_T$ in the left panel of Fig.~\ref{fig:nT_limit} can be translated into those on $R_\sigma$, which are depicted in the right panel of Fig.~\ref{fig:nT_limit}.
The right panel shows that $R_\sigma$ is constrained to $R_\sigma \lesssim {\cal O}(10)$ when $r \gtrsim 0.02$. 
We stress that the upper bound on $R_\sigma$ is obtained once the limit on $n_T$ becomes available.

\begin{figure}[htbp]
  \begin{center}
 \resizebox{140mm}{!}{\includegraphics{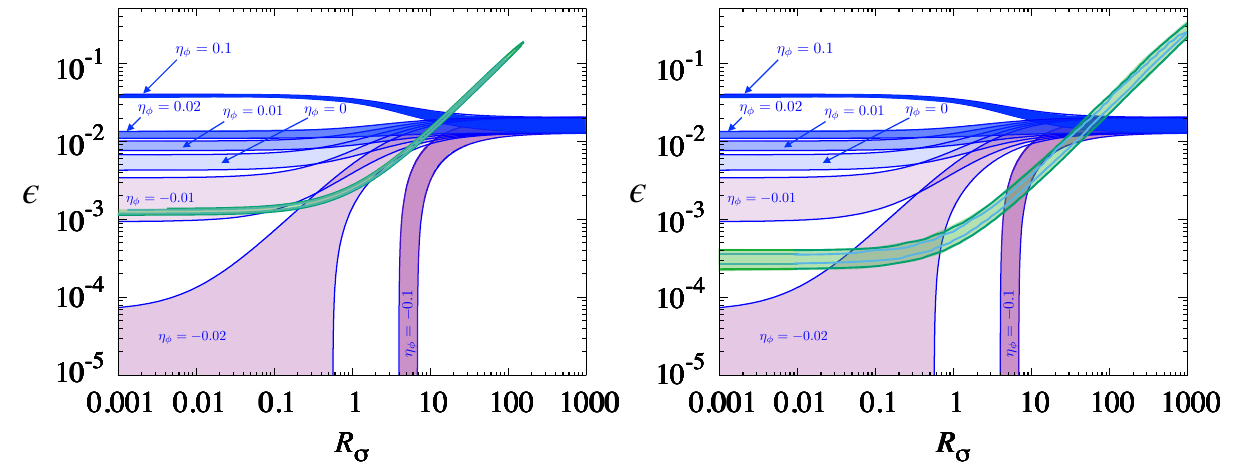}}\\
  \resizebox{140mm}{!}{\includegraphics{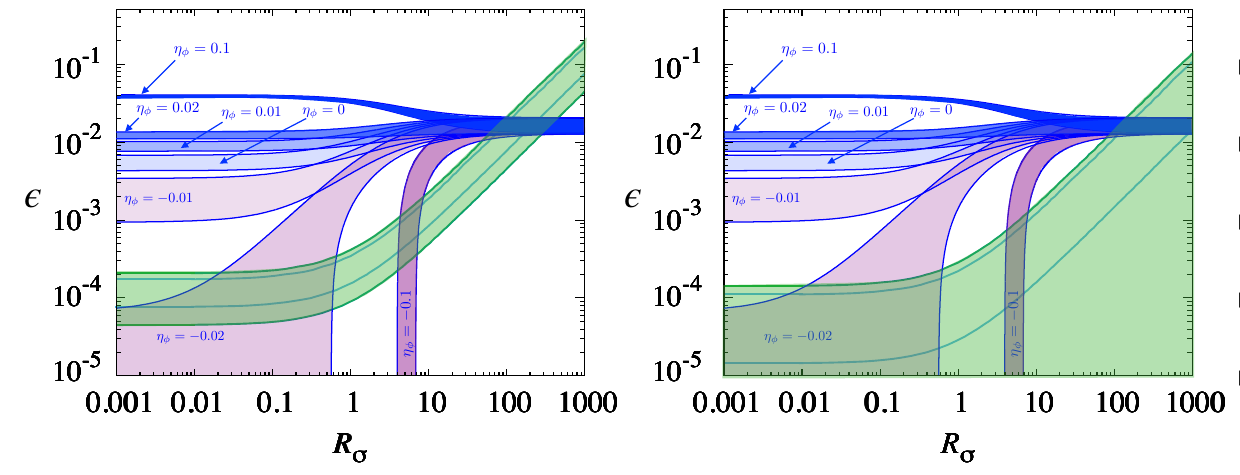}}
  \end{center}
  \caption{ \label{fig:Rsig_r}
  1$\sigma$ and 2$\sigma$ expected constraints on the $R_\sigma$--$\epsilon$ plane from LiteBIRD (green).
  The fiducial values are taken as $(r, n_T) = (0.02, -0.0025)$ (top left), $(0.005 -0.000625)$ (top right), $(0.002, -0.00025)$ (bottom left), and $(0.001, -0.000125)$ (bottom right) that satisfy the single-field consistency relation \eqref{eq:consistency_inf}. 
  Constraints on $n_s$ from Planck $(n_s = 0.967 \pm 0.0074 ~(95 \% \,{\rm CL}))$ are also shown in blue/magenta for the cases of $ \eta_\phi =-0.1, -0.02, -0.01, 0, 0.01, 0.02$, and $0.1$.
  Here we assume the mass of the spectator to be light enough such that $\eta_\sigma = 0$ is a good approximation.
  }
\end{figure}

In Fig.~\ref{fig:Rsig_r}, we also show the expected constraints from LiteBIRD in the $R_\sigma$--$\epsilon$ plane.
We assume the fiducial model to be $ (r, n_T) = (0.02, -0.0025)$ (top left), $(0.005 -0.000625)$ (top right), $(0.002, -0.00025)$ (bottom left), and $(0.001, -0.000125)$ (bottom right).
These points satisfy the single-field consistency relation \eqref{eq:consistency_inf} between $r$ and $n_T$.
We also display the parameter region that gives the scalar spectral index in the 2$\sigma$ allowed range from Planck \cite{Planck:2018jri}, $0.9596 < n_s < 0.9744$, for several fixed values of $\eta_\phi$.  
In this figure, we assume that the spectator field is light enough such that its mass is negligibly smaller than the Hubble rate during inflation and thus we set $\eta_\sigma = 0$.
The spectral index $n_s$ is calculated from Eq.~\eqref{eq:ns_mixed}, and hence, with $\eta_\sigma =0$, we need to fix $\eta_\phi$ in order to predict $n_s$ in the $R_\sigma$--$\epsilon$ plane.
In the large $R_\sigma$ limit, the spectral index becomes $n_s \simeq 1 - 2\epsilon$ independently of the value of $\eta_\phi$, and hence the region $0.9596 < n_s < 0.9744$ converges to the band $0.0128 \lesssim \epsilon \lesssim 0.020$ in this limit.
On the other hand, in the single-field limit $R_\sigma \ll 1$ the scalar spectral index is given by the formula $n_s = 1 - 6 \epsilon + 2\eta_\phi$, and in particular, it reduces to $n_s \simeq 1 + 2\eta_\phi$ for $\epsilon \ll |\eta|$.
In such a region, the value of $\eta_\phi$ must be in the range of $- 0.020 < \eta_\phi < - 0.0128$ to match the Planck constraints on $n_s$.

Having the above considerations in mind, we now discuss the behavior of each panel in Fig.~\ref{fig:Rsig_r}.
When the fiducial value of $r$ is relatively large as $r = 0.02$ (the top left panel), 
LiteBIRD gives an upper limit on $\epsilon$ since $n_T$ is constrained somewhat stringently. 
More importantly, given the current constraints on $n_s$ (shown with blue/magenta for each value of $\eta_\phi$), the spectator field cannot give a dominant contribution.
As seen from the top left panel, the range $R_\sigma \gtrsim 10$ is not allowed once LiteBIRD and Planck constraints are both taken into account.
This means that most parameter range of this multi-field model would be excluded once the tensor-to-scalar ratio $r \sim 0.02$ is detected with LiteBIRD. 
The tendency is similar for $r = 0.005$ (top right) and $r = 0.001$ (bottom left), though larger $R_\sigma$ are now allowed for some limited range of $\epsilon$.
When the fiducial tensor-to-scalar ratio decreases to $r = 0.001$, we only get an upper limit on $r$ as shown in the bottom right panel in Fig.~\ref{fig:nt_r_single}.
In this case, multi-field inflation models with $R_\sigma \gtrsim 100$ are favored when $\eta_\phi \gtrsim -0.02$. Although above statements somewhat depend on the assumption of the inflaton potential (i.e., $\epsilon$ and $\eta_\phi$), the arguments here indicate that LiteBIRD can test multi-field inflation models, particularly when the information of the scalar spectral index is combined.

\subsection{Case (ii): a multi-field fiducial model}

\begin{figure}[htbp]
  \begin{center}
  \resizebox{170mm}{!}{\includegraphics{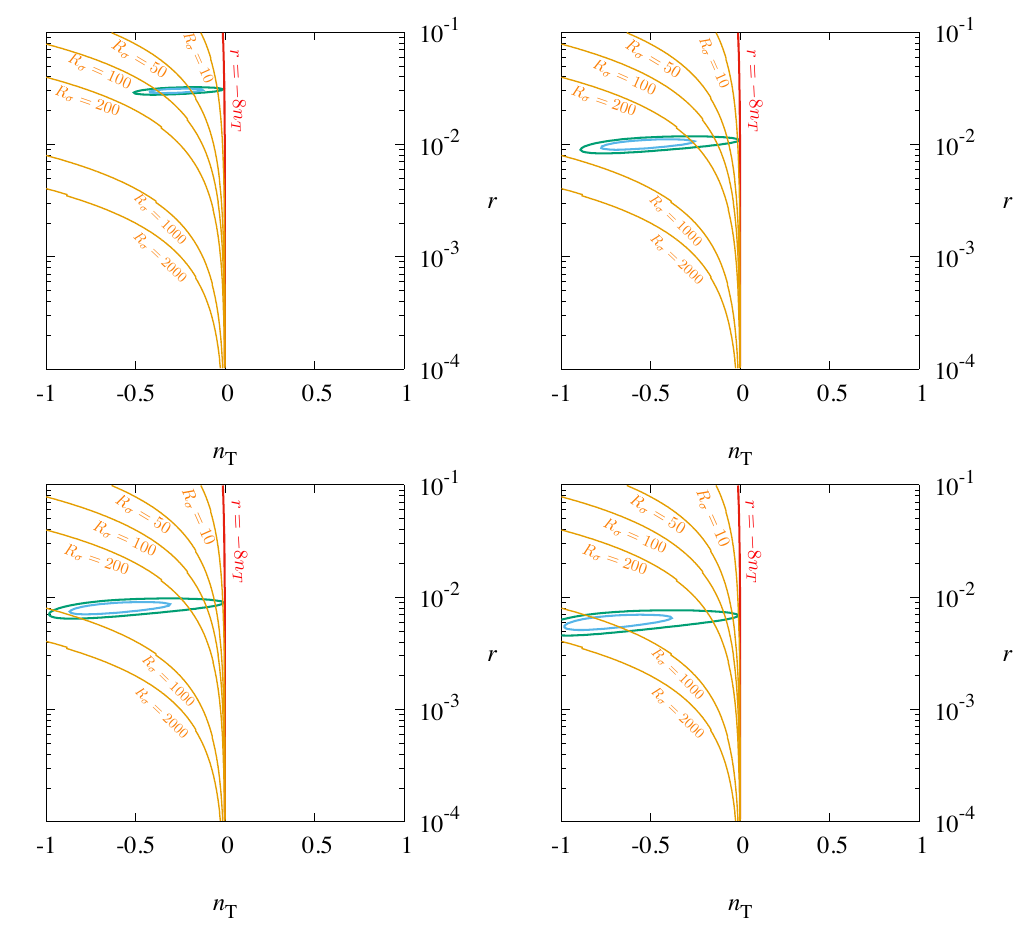}}
  \end{center}
  \caption{\label{fig:nt_r_multi}
  1$\sigma$ and 2$\sigma$ expected constraints on the $n_T$--$r$ plane from LiteBIRD.
  The fiducial values are taken as $(r,n_T) = (0.03, -0.28)$ (top left), $(0.01, -0.56)$ (top right), $(0.008, -0.64)$ (bottom left), and $(0.006, -0.75)$ (bottom right).
  These fiducial values are not from the single-field inflation consistency relation \eqref{eq:consistency_inf} but rather from \eqref{eq:consistency_mixed} with $R_\sigma \simeq 74, 450, 640$, and $1000$, respectively.
  }
\end{figure}

Now we present the results with the fiducial model taken to be multi-field models.
This means that the single-field consistency relation does not hold anymore, but instead the multi-field counterpart \eqref{eq:consistency_mixed} is satisfied at the fiducial point.
In Fig.~\ref{fig:nt_r_multi}, we show the expected constraints from LiteBIRD for the fiducial model of $(r,n_T) = (0.03, -0.28)$ (top left),  $(0.01, -0.56)$ (top right), $(0.008, -0.64)$ (bottom left), and $(0.006, -0.75)$ (bottom right), which correspond to $R_\sigma \simeq 74, 450, 640$ and $1000$, respectively.
As seen from the figure, the single-field model can be excluded at almost 2$\sigma$ level when the tensor spectral index at the fiducial point is negatively as large as the value taken in each panel.

Models with a negatively large value of $n_T$ correspond to those with relatively large $\epsilon$.
Inflation models with large $\epsilon$ has not been considered much in the literature, since they cannot be consistent with observational constraints as a single-field model.
However, in the multi-field framework we can construct such models, and we discuss an explicit example in Appendix~\ref{app:example}.

\section{Conclusion} \label{sec:conclusion}

In this paper we investigated the expected constraints on the tensor-to-scalar ratio $r$ and the tensor spectral index $n_T$ from the future CMB $B$-mode polarization experiment LiteBIRD.
We first summarized the differences between the  predictions for single-field inflation and multi-field inflation in Sec.~\ref{sec:multi_inf}, and then discuss the method used for our likelihood analysis and assumptions about the foreground in Sec.~\ref{sec:method}.
We introduced a parameter that represents the fraction of the residual foreground, and adjust its value in such a way that it reproduces the expected constraints from LiteBIRD provided in the literature \cite{LiteBIRD:2022cnt} at the fiducial point $n_T = 0$.
This allows us a simple and quicker parameter scan while still putting reasonable constraints.

Using this machinery, we discussed to what extent the future LiteBIRD experiment can test multi-field inflation models.
First we assumed single-field inflation as the fiducial model and obtain the expected constraints.
As shown in Fig.~\ref{fig:nt_r_single}, while it would be very challenging to test the consistency relation \eqref{eq:consistency_inf} that holds in single-field inflation models, multi-field models can be tested once the results from LiteBIRD is available in combination with the information on the scalar spectral index $n_s$ that has been severely constrained by Planck.
In particular, when the tensor-to-scalar ratio is detected with LiteBIRD, the parameter space for multi-field models with a light spectator field is identified or stringently constrained, which may be helpful in constructing  the inflation model as realized in nature from the viewpoint of high energy physics.
Interestingly, even if LiteBIRD put only an upper bound on $r$, such a constraint still gives useful information on multi-field models.
Depending on the inflaton model, the upper bound on $r$ can be translated into the lower bound on $R_\sigma$, the fraction of the spectator field to the curvature perturbation.
Such a lower bound, if obtained, would suggest a preference for multi-field models.

If the absolute value of the tensor spectral index is relatively large, we can test the so-called consistency relation $r = - 8 n_T$ that holds in the single-field inflation model.
We argue that such a test is possible when the tensor spectral index is negatively large as $n_T \sim - {\rm O}(0.1)$.
We also discuss that, even with such a negatively large $n_T$, inflation models can be constructed in the multi-field framework and that they can be tested with LiteBIRD.

As argued in this paper, the information on the tensor power spectrum such as $r$ and $n_T$ would be useful to test multi-field inflation models even when constraints are available only partially.
Although we discussed only multi-field models in this paper, it would be interesting to investigate how the LiteBIRD experiment can be used to test various inflation models, which we leave for future work.

\section*{Acknowledgements}
This work was supported by JSPS KAKENHI Grant Numbers 23K19048 (RJ), 22H01215 (TM), 19K03874 (TT) and MEXT KAKENHI 23H04515 (TT). 
Members of QUP, KEK were supported by the World Premier International Research Center Initiative (WPI) of MEXT.

\pagebreak

\appendix
\noindent
{\LARGE \bf Appendix} 

\section{Foreground \label{app:foreground} }

In this appendix, we summarize the foreground power spectra adopted in our analysis.  Power spectra from synchrotron emission and thermal dust emission can be parametrized as
\begin{equation}
C_{ l}^{BB, {\rm syn}} (\nu_i) = A^{\rm syn} 
\left( \frac{l}{l_{0 s}} \right)^{\alpha_s -2}
\left( \frac{\nu_i}{\nu_{0 s}} \right)^{2\beta_s}  , 
\end{equation}
and 
\begin{equation}
C_l^{BB, {\rm dust}} (\nu_i) 
= 
p^2 A^{\rm dust} 
 \left( \frac{l}{l_{0 D}} \right)^{\alpha_D-2} 
\left( \frac{\nu_i}{\nu_{0 D}} \right)^{2\beta_D+2} 
\left( \frac{e^{h\nu_{0D} / k_B T_D} -1}{e^{h\nu_i / k_B T_D} -1} \right)^2,
\end{equation}
where $A^{\rm syn}$ and $A^{\rm dust}$ represent the amplitude of the synchrotron and dust foreground, $\alpha_s (\alpha_D)$ and $\beta_s (\beta_D)$ characterize $l$ and  frequency dependence of synchrotron (dust), and $p$ is the dust polarization fraction.
The values of these parameters adopted in the analysis are summarized in Table~\ref{tab:param_foreground}.
These values are chosen to be consistent with those given by Planck \cite{Planck:2018yye,Planck:2018fzr,Planck:2018gnk}.

\begin{table}[hb]
  \centering
  \begin{tabular}{l | l}
\hline \hline 
Synchrotron & ~~~Dust~~~  \\  \hline
$\alpha_s = -0.76 $ & $\alpha_D = -0.5$ \\ \\
$\beta_s = -3.1 $ & $\beta_D = 1.55$ \\ \\
$ \nu_{\rm 0 s} =30 {\rm GHz} $  &$ \nu_{\rm 0 D} = 353 {\rm GHz}$ \\ \\
$ l_{\rm 0 s} = 80 $  &$ l_{\rm 0 D} =80 $ \\ \\
$ A^{\rm syn} = 7.85 \times 10^{-4}~(\mu{\rm K})^2 $  &  $ A^{\rm dust} = 0.2~(\mu{\rm K})^2 $  \\ \\
  &  $T_D = 19.6~{\rm K}$  \\ \\
  &  $p=0.1$  \\
\hline
\end{tabular} 
  \caption{ \label{tab:param_foreground} Parameters for the foreground} 
\end{table}

The noise power spectrum of the foreground template map for a frequency channel $\nu_i$ is given by
\begin{equation}
N_{l}^{{BB, \rm FG}} (\nu_i)= \frac{N^{({\rm tot})BB} _{l {\rm (inst)}}}{N_{\rm ch} ( N_{\rm ch}-1)/4} 
\left[  \left( \frac{\nu_i}{\nu_{s, {\rm ref}}} \right)^{2\beta_s} + \left( \frac{\nu_i}{\nu_{D, {\rm ref}}} \right)^{2\beta_D+2} \right] \,, 
\end{equation}
where $N^{({\rm tot})BB} _{l {\rm (inst)}}$ is the instrumental power spectrum for the total  channel, $N_{\rm ch}$ is the number of the channels. $\nu_{s, {\rm ref}}$ and $\nu_{D, {\rm ref}}$ are taken to be the lowest and highest frequencies included in the analysis.

\section{An example model with large negative $n_T$\label{app:example}}

As an example of an inflationary model that realizes large negative value of $n_T$ while satisfying observational constraints, we consider a three-field setup. The scalar potential is given by the following simple form:
\begin{equation}
\label{eq:three_field_V}
V(\phi, \sigma, \chi) = \frac14 \lambda_\phi \phi^4 + \frac12 m_\sigma^2 \sigma^2  + \frac12 m_\chi^2 \chi^2 \,,
\end{equation}
where $\lambda_\phi$ is the quartic coupling for $\phi$, while $m_\sigma$ and $m_\chi$ are the masses for $\sigma$ and $\chi$, respectively.
Here $\phi$ plays the role of the inflaton, $\sigma$ is the curvaton, and $\chi$ is a scalar field causing the second inflationary phase after the curvaton-dominated epoch.
We consider the case where the curvaton gives a dominant contribution to the primordial power spectrum $R_\sigma \gg 1$, which can be realized with a small initial amplitude of the curvaton $\sigma_\ast \lesssim 0.1 \, M_{\rm pl}$ (see Eq.~\eqref{eq:R_sigma}).
In order for the second inflaton to be caused by $\chi$, its initial value is taken to be $ \chi_\ast \simeq 10 \, M_{\rm pl}$. 

The curvature perturbation generated from these fields are evaluated as 
\begin{equation}
 \zeta^{(\phi)}  = \frac{1}{\sqrt{2 \epsilon M_{\rm pl}^2}} \delta \phi_\ast  \, 
 \qquad
 \zeta^{(\sigma)}  = \frac{2 r_{\rm dec}}{3 \sigma_\ast} \delta \sigma_\ast  \, 
\qquad
 \zeta^{(\chi)}  =  \frac{\chi_\ast}{2  M_{\rm pl}^2} \delta \chi_\ast  \,.
\end{equation}
The curvature perturbations from $\phi$ and $\sigma$ are given by the standard formulas for the inflaton and curvaton, respectively (see, e.g., \cite{Sasaki:2006kq,Enqvist:2013paa}), while that for $\chi$ can be evaluated in the same manner as the inflating curvaton scenario \cite{Langlois:2004nn,Moroi:2005kz,Ichikawa:2008iq,Dimopoulos:2011gb} (see also \cite{Enqvist:2019jkb} for a similar setup).
Since the initial value of $\chi$ is assumed to be large, the contribution from $\chi$ for $\zeta$ is almost the same as that from $\phi$:
\begin{equation}
\frac{ \zeta^{(\chi)}}{ \zeta^{(\phi)}}  = \sqrt{2 \epsilon} \frac{\chi_\ast}{M_{\rm pl}} \,.
\end{equation}
However, we can safely ignore the contribution from $\chi$ since the curvaton generates much larger primordial density fluctuations.

The slow-roll parameters in this model are given by
\begin{equation}
\epsilon = \frac{1}{N} \,, 
\qquad
\eta_\phi = \frac{3}{2N} \,,
\qquad
\eta_\sigma = \frac{m_\sigma^2}{3H^2} \,,
\end{equation}
with $N$ being the number of $e$-fold at the time when the reference scale exited the horizon. When the spectator gives the dominant contribution to the power spectrum, the scalar spectral index is given by
\begin{equation} 
n_s  -1 =  -2 \epsilon + 2 \eta_\sigma \,.
\end{equation}
Since $n_s - 1  \simeq -0.04$ from the Planck measurement, we need $\epsilon \simeq 0.02 + \eta_\sigma$.
Therefore, for a relatively large value for $m_\sigma$, we have $\epsilon \simeq 0.05$ and hence $n_T \simeq -0.1$.
Since $\epsilon$ is related to the number of $e$-folds as $\epsilon = 1/N $, we need $N \simeq 20$, which is possible if the second inflationary phase due to the $\chi$ field lasts for $e$-folds of $\sim 30$.
We note that, though the slow-roll parameter in this scenario is relatively large as $\epsilon = 0.05$, the running parameter is still small as $\alpha_s \sim {\cal O}(10^{-3})$ and is well within the observational bound from Planck \cite{Planck:2018jri}.

\clearpage 
\bibliography{CMB_inf}

\end{document}